\documentclass[11pt,epsf]{iopart}

\usepackage{iopams}
\usepackage{graphics}
\usepackage{graphicx}

\usepackage{epsfig}
\newcommand{\dd}{\hbox{d}}

\newcommand{\ee}{\hbox{e}}

\newcommand{\p}{\partial}

\newcommand{\eps}{\varepsilon}

\begin{document}

\title[]{Energy flux distribution in a two-temperature Ising model}

\author{Vivien Lecomte$^{(1)}$, Zolt\'an R\'acz$^{(2)}$ and Fr\'ed\'eric van Wijland$^{(1,3)}$}

\address{\ $^{(1)}$Laboratoire de Physique Th\'eorique, CNRS UMR 8627,
Universit\'e de Paris-Sud, 91405 Orsay cedex, France}

\address{\ $^{(2)}$Institute for Theoretical Physics, HAS Research Group,
E\"otv\"os University, 1117 Budapest, P\'azm\'any s\'et\'any 1/a, Hungary}

\address{\ $^{(3)}$P\^ole Mati\`ere et Syst\`emes Complexes, CNRS UMR 7057, 
Universit\'e Denis Diderot (VII), 4 place Jussieu, 75005 Paris, France}

\begin{abstract}
The nonequilibrium steady state of an infinite-range Ising model
is studied. The steady state is obtained by dividing the spins
into two groups and attaching them to two heat baths generating
spin flips at different temperatures. In the thermodynamic limit,
the resulting dynamics can be solved exactly, and the
probability flow in the phase space can be visualized. We
can calculate the steady state fluctuations
far from equilibrium and,
in particular, we find the exact probability distribution of the
energy current in both the high- and low-temperature phase.


\end{abstract}

\pacs{05.50.+q, 05.40.-a, 05.70.Ln, 05.60.-k}


\newpage

\section{Introduction}

The infinite range Ising model, in which each individual spin  interacts with
the remaining $N-1$ ones  has served as a useful
testbench for many ideas in various subfields of statistical physics, ranging from critical dynamics to
spin glasses. The reason is twofold: it is relatively easy to come up with exact
yet nontrivial results for this system (in the large system limit at least), while at the same
time it stands for a genuine interacting system which further possesses a phase
transition from a disordered into an ordered state as the temperature is
lowered. In the present work we will follow Ruijgrok and
Tjon~\cite{ruijgroktjon} who, by endowing the system with spin-flip dynamics, provided the first example of an exactly solvable
critical dynamics problem back in 1973. In the meantime the subjects of interest
have drifted towards other issues, but the technical motivations for using the infinite
range Ising model remain. Our physical motivations have their roots in the study and
identification of the generic
properties of nonequilibrium steady-states (NESS).\\

Among the very few exact statements that can be made on NESS, the
recent fluctuation --or Gallavotti--Cohen--
theorem~\cite{gallavotticohen} plays a prominent role. This theorem is
a symmetry property of the entropy current distribution function. As
such it provides, at least formally, a prescription for obtaining an
infinite set of Green--Kubo like relations connecting current
fluctuations to the system's response upon external forcing. While
establishing a Gallavotti--Cohen theorem for Markov processes can be
achieved in general terms~\cite{kurchan,lebowitzspohn}, explicit
computations of current distribution functions are rare. For systems
{\it close} to equilibrium such as boundary-driven lattice gases,
Bodineau and Derrida~\cite{bodineauderrida}, and Bertini {\it et
al.}~\cite{bertinidesolegabriellijonalasiniolandim2} have provided a
general method for computing current distribution functions from the
sole knowledge of the Onsager response coefficients. However for the
generic case of systems maintained in a steady-state {\it far} from
equilibrium, no general principle has hitherto been unraveled. To the
best of our knowledge a single exact calculation exists for the
totally asymmetric exclusion process, a degenerate case for which the
Gallavotti--Cohen theorem (in the version of Lebowitz and
Spohn~\cite{lebowitzspohn}) does not hold. For this reason we have
sought to exhibit a system driven far from equilibrium for which
--albeit mean-field-- such a current distribution function is
accessible: an Ising model in which spins are connected to heat baths
at different temperatures.\\

There are several ways of driving the infinite range Ising model  into a NESS.
One that is inspired from a series of recent
works~\cite{raczzia,schmittmannschmuser,hilhorstschmittmannzia,garridomarro}
consists in
coupling spins to independent heat baths, thus creating a macroscopic energy current by means of a
bulk drive. In the particular version we have coined, $N$ Ising spins $\sigma_i$ have a
ferromagnetic interaction energy
\begin{equation}
{\cal H}[\{\sigma_i\}]=-\frac{1}{2N }\left(\sum_i\sigma_i\right)^2
\end{equation}
and the dynamics of $N/2$ spins is generated by a heat bath of inverse
temperature $\beta_1$ while the dynamics of the remaining half
of the spins is driven by another heat bath at a different inverse  temperature $\beta_2$. This
state of affairs, namely the existence of two heat baths at unequal temperatures, leads to a steady
energy current flowing through the spins
from the warmer bath towards the colder one.\\

The physical results we have obtained are concerned with the steady-state measure and the
energy current distribution function. We have been able to provide an exact
solution to the Fokker-Planck equation governing the probability distribution
of magnetization fluctuations, thus leading to the first example of an $N$-body
nonequilibrium system for which probability
flow lines in phase space can actually be visualized. Our second achievement is
to have  provided the large deviation function of the energy current. The methods we
have resorted to rest on the various formulations of the master equation
governing the microscopic dynamics. On one hand, by means of a Van Kampen~\cite{vankampen}
expansion of the magnetizations around their mean values, we have obtained
solvable Fokker-Planck equation describing steady-state fluctuations. Technically, this amounts to finding an eigenfunction of
the master equation evolution operator. On the other hand, by means of a mapping
of the master equation onto a quantum hamiltonian, we have determined the energy
current distribution function, which, in technical terms, has proved to be an
eigenvalue problem.\\

Our first task  will be to provide an accurate description of the stationary state distribution
(Sec.~3), which we will present after having precisely characterized our model
and its steady-state properties (Sec.~2). Sec.~4 will be devoted to a complete study of the energy current distribution
function, in the light of the Gallavotti--Cohen theorem. Our concluding remarks will be followed by
two appendices generalizing the aforementioned results to an infinite-range Ising model in contact with
more than two heat baths. 

\section{A two temperature Ising model: phase diagram and steady-state properties}
\subsection{Microscopic dynamics}
The energy of a configuration of $N$ Ising spins $\bsigma=\{\sigma_i\}$ is given by
\begin{equation}
{\cal H}[\bsigma]=-\frac{M^{2}}{2N},\;M=\sum_{i=1}^N\sigma_i
\end{equation}
We now divide the $N$ spins into two groups with labels $1$ and $2$ of $N/2$ spins each. 
A spin $\sigma_j$ from set 1 flips with a rate
\begin{equation}\label{rateE}
\forall j \in 1,\;w_1(\sigma_j\to-\sigma_j)=\ee^{-\beta_1\sigma_j M/N}
\end{equation}
Spins from group 1  try to equilibrate at inverse temperature $\beta_1$ with respect to ${\cal H}$.
Similarly, a spin from group 2 flips according to
\begin{equation}\label{rateO}
\forall j \in 2,\;w_2(\sigma_j\to-\sigma_j)=\ee^{-\beta_2\sigma_j M/N}
\end{equation}
This is the infinite-range counterpart to the one-dimensional systems considered by R\'acz and
Zia~\cite{raczzia}, and Schmittmann  and Schmuser~\cite{schmittmannschmuser}. 
Denoting by
\begin{equation}\label{defb1b2}
\beta= \frac{\beta_1+\beta_2}{2 },\;\eps=\frac{\beta_1-\beta_2}{2 }
\end{equation}
we see that when the temperatures are equal, $\beta_1=\beta_2=\beta$, or $\eps=0$, the system reaches equilibrium at temperature
$\beta$. This is because the rates (\ref{rateE},\ref{rateO}) then satisfy detailed balance with respect
to the Gibbs distribution $Z^{-1 }\ee^{-\beta {\cal H}}$. Though the precise expressions of the
rates we have chosen differ from the original Glauber rates, they possess the same qualitative
properties with some advantages in the large-system limit
discovered by Ruijgrok and Tjon~\cite{ruijgroktjon}.\\

\subsection{Phase diagram}
Introducing the mean magnetizations
\begin{equation}
m_1=\frac{1}{N}\langle \sum_{j\in 1}\sigma_j\rangle,\;\;m_2=\frac{1}{N}\langle \sum_{j\in
2}\sigma_j\rangle,\;\;m=m_1+m_2
\end{equation}
we may
find the following evolution equations for the averages:
\begin{equation}\label{autocoh}
\frac{\dd m_1}{\dd t}=-2m_1\cosh\beta_1 m+\sinh\beta_1 m,\;\frac{\dd m_2}{\dd
t}=-2m_2\cosh\beta_2 m+\sinh\beta_2 m
\end{equation}
from which one deduces that in the steady-state (provided it exists)
\begin{equation}\label{eqpourm}
m=\frac 12\left(\tanh\beta_1 m+\tanh\beta_2 m\right)
\end{equation}
Interestingly, although the transition rates (\ref{rateE},\ref{rateO})
are different from the standard Glauber rates, they lead to the same
steady-state average magnetization.
From (\ref{eqpourm}) we deduce that in the steady-state the system
undergoes a second order phase transition from a high-temperature
disordered state at $\beta< 1$ in which $m_1=m_2=m=0$ to a low-temperature ordered (doubly
degenerate) state at
$\beta>1$ with nonzero magnetizations. In the $\beta \to 1^+$ limit at $\eps$ fixed one finds
\begin{equation}
\!\!\!\!\!\!\!\!\!\!\!\!\!\!\!\!\!\!\!\!\!\!\!\!\!\!\!m\simeq\pm\frac{\sqrt{3}}{\sqrt{1+3\eps^2}}\sqrt{\beta-1},\;m_1\simeq\pm\frac{1+\eps}{2}\frac{\sqrt{3}}{\sqrt{1+3\eps^2}}\sqrt{\beta-1},\;m_2\simeq\pm\frac{1-\eps}{2}\frac{\sqrt{3}}{\sqrt{1+3\eps^2}}\sqrt{\beta-1}
\end{equation}
According to the magnitude of the nonequilibrium drive $\eps$ it may be seen that the ordered state may be either
ferromagnetic ($|\eps|<1$) or antiferromagnetic ($|\eps|>1$, if one allows for
negative temperatures, as we shall discuss in our conclusion in Sec.~\ref{finalcomments}).
\subsection{Entropy and energy currents}
Following the prescription of Lebowitz and Spohn~\cite{lebowitzspohn} we may define a time
integrated instantaneous entropy current by
\begin{equation}\label{entropycurrent}
{\cal Q}_S(t)=
\ln\frac{W(\bsigma^{(0)}\to\bsigma^{(1)})}{W(\bsigma^{(1)}\to\bsigma^{(0)})}
...\frac{W(\bsigma^{(k-1)}\to\bsigma^{(k)})}{W(\bsigma^{(k)}\to\bsigma^{(k-1)})}
\end{equation}
where $\bsigma^{(0)}=\bsigma(0),...,\bsigma^{(k)}=\bsigma(t)$ is the sequence of states occupied by the
system over the time interval $[0,t]$ (this is the history of the system between $0$ and $t$). The
rates $W(\bsigma\to\bsigma')$ of hopping from configuration $\bsigma$ to configuration $\bsigma'$ between $t$ and $t+\dd t$
are easily deduced from (\ref{rateE},\ref{rateO}). Inserting the explicit expressions for the
$W(\bsigma\to\bsigma')$ leads to
\begin{equation}\label{entropycurrentdecomposition}
{\cal Q}_S(t)=
-\beta\left({\cal H}[\bsigma(t)]-{\cal H}[\bsigma(0)]\right)+\eps Q(t)
\end{equation}
where we identify $Q$ as the integrated energy current:
\begin{equation}\label{energycurrentdef}
Q(t)=-\frac{2}{N}\sum_{n=0}^k(\pm)\sigma_{j_n}(M_n-\sigma_{j_n})
\end{equation}
where $\sigma_{j_n}$ is the spin being flipped at time $n$ and $M_n$ is the total magnetization at
that moment. The sign $+$ (resp. $-$) corresponds to flipping a spin from group 1 (resp. 2). Note
that ${\cal H}[\bsigma(t)]-{\cal H}[\bsigma(0)]$ being bounded over time, ${\cal Q}_S(t)$ and $\eps Q(t)$ have the same large deviation functions. It is clear that on average,
\begin{equation}\label{J1J2}
J_\eps=\frac{\langle Q(t)\rangle}{t}=-\frac{2}{N}\langle\sum_{j\in 1}\sigma_j (M-\sigma_j)\ee^{-\beta_1\sigma_j
M/N}-\sum_{j\in 2}\sigma_j (M-\sigma_j)\ee^{-\beta_2\sigma_j M/N}\rangle
\end{equation}
While interpreting ${\cal Q}_S(t)$ as an integrated entropy current requires an elaborate
reasoning~\cite{lebowitzspohn}, the physical meaning of $J_\eps$ as an  energy current is much more
intuitive.  Indeed, the total energy of the system is constant on average in the
steady-state:
\begin{equation}
\frac{\dd\langle{\cal H}\rangle}{\dd t}=0=-(J_1+J_2)
\end{equation}
where $J_\alpha$ is the energy flux due to spin-flips caused by heat-bath
$\alpha$ (for instance $J_1$ is the first term on the rhs of (\ref{J1J2})). The
quantity $J_\eps=J_1-J_2$ is therefore a measure of the energy flowing from group 1
towards group 2. Hence the related entropy current $J_S$ must read 
\begin{equation}
J_S=\beta_1 J_1+\beta_2 J_2=\eps J_\eps
\end{equation}
This interpretation of $\eps J_\eps$ as an entropy current has been discussed, on the grounds of
phenomenological thermodynamics, by R\'acz and Zia~\cite{raczzia}.

As described in appendix B, there is no immediate link between the entropy current and
an energy current for a system in contact with more that two heat baths.

\section{Stationary state distribution}
\subsection{Van Kampen expansion and Fokker-Planck equation}
In this section we derive a Fokker-Planck equation governing the probabililty
$P(x_1,x_2,t)$ of observing the following  fluctuations of the spin magnetizations:
\begin{equation}
x_\alpha=\frac{\sum_{j\in \alpha}\sigma_j-N m_\alpha}{\sqrt{N}},\;\alpha=1,2
\end{equation}
This is the Van Kampen~\cite{vankampen} expansion of the master equation around the mean
magnetizations $m_\alpha$. The $\sqrt{N }$ rescaling is precisely designed for the $x_\alpha$ to
have order 1 fluctuations. We find that $P(x_1,x_2,t)$ satisfies the following Fokker-Planck equation:
\begin{equation}
\p_t P=-\p_{x_1} {\cal J}_1-\p_{x_2} {\cal J}_2
\end{equation}
where the probability current is given by
\begin{equation}
{\cal J}_\alpha= f_\alpha(x_1,x_2)P- D_\alpha\p_{x_\alpha} P
\end{equation}
The two-dimensional force $(f_1,f_2)$ does not derive from a potential unless both heat baths are at the same temperature. The general expression of the
force components is
\begin{eqnarray}
f_1(x_1,x_2)=((\beta_1-2)x_1+\beta_1 x_2)\cosh\beta_1 m-2\beta_1 (x_1+x_2)m_1\sinh\beta_1 m\nonumber\\
f_2(x_1,x_2)=((\beta_2-2)x_2+\beta_2 x_1)\cosh\beta_2 m-2\beta_2 (x_1+x_2)m_2\sinh\beta_2 m
\end{eqnarray}
The diffusion constants are given by
\begin{equation}
D_\alpha=\cosh\beta_\alpha m-2 m_\alpha\sinh\beta_\alpha m=\sqrt{1-4 m_\alpha^2}
\end{equation}
In the high temperature phase, using that $\beta_{1/2}=\beta\pm \eps$, this may easily be cast in the following form
\begin{equation}
f_1(x_1,x_2)=-\p_{x_1} U_\eps+\eps x_2,\;f_2(x_1,x_2)=-\p_{x_2} U_\eps-\eps x_1
\end{equation}
where the potential energy has the expression
\begin{equation}
U_\eps(x_1,x_2)=(1-\beta)\frac{(x_1+x_2)^2}{2}+\frac{(x_1-x_2)^2}{2}-\eps\frac{x_1^2-x_2^2}{2}
\end{equation}
In the high temperature phase, to which the ensuing analysis will be confined for simplicity (a
general solution is provided in appendix A), where $m_1=m_2=0$, we thus have to solve
\begin{equation} 
\p_t P=0=-\p_{x_1} {\cal J}_1-\p_{x_2} {\cal J}_2
\end{equation}
where the probability current reduces to
\begin{equation}
{\cal J}_1=((\beta-2)x_1+\beta x_2+\eps(x_1+x_2))P- \p_{x_1} P ,\;
{\cal J}_2=((\beta-2)x_2+\beta x_1-\eps(x_1+x_2))P- \p_{x_2} P
\end{equation}
When $\beta_1=\beta_2=\beta$, that is in equilibrium, the distribution reads
\begin{equation}
P(x_1,x_2)\sim \exp\left[-(1-\beta)\frac{(x_1+x_2)^2}{2}-\frac{(x_1-x_2)^2}{2}\right]
\end{equation}
We may find the exact solution to the
Fokker-Planck equation by having the intuition, following  \cite{vankampen},
that the effective potential $U_{\hbox{\tiny eff}}$ defined by
\begin{equation}\label{FP1}
P(x_1,x_2)=Z^{-1} \exp(-U_{\hbox{\tiny eff}})
\end{equation}
will be quadratic in terms of $x_1$ and $x_2$. This is suggested by the force being linear
in $x_1$ and
$x_2$. And indeed this naive assumption leads to the effective potential
$U_{\hbox{\tiny eff}}$ given by 
\begin{equation}
U_{\hbox{\tiny eff}}(x_1,x_2)=\frac{1-\beta}{2}
(x_1+x_2)^2+\frac{2}{4+J^2}\left(x_1[1-J/2]-x_2[1+J/2]\right)^2
\end{equation}
where we have introduced the constant $J\equiv\frac{2\eps}{2-\beta}$.
It is then an easy task to compute the mean energy current $J_\eps$:
\begin{eqnarray}
J_\eps=&\langle\sum_{j\in 1}(-2\sigma_j^z
(M^z-\sigma_j^z)/N)\ee^{-\beta_1 \sigma_j^z M^z/N}-\sum_{j\in
2}(-2\sigma_j^z (M^z-\sigma_j^z)/N)\ee^{-\beta_2 \sigma_j^z
M^z/N}\rangle\nonumber\\
\;\;\;\;=&2\eps\langle(x_1+x_2)^2\rangle-2\langle(x_1^2-x_2^2)\rangle\nonumber\\
\;\;\;\;=&\frac{2\eps}{2-\beta}=J
\end{eqnarray}
In terms of the magnetization fluctuations the solution to the
Fokker-Planck equation reads
\begin{equation}
\!\!\!\!\!\!\!\!P(x_1,x_2)\sim\exp\left[-\frac{1-\beta}{2}(x_1+x_2)^2-\frac{2}{4+J^2}\left(x_1[1-J/2]-x_2[1+J/2]\right)^2\right]
\end{equation}
A few comments are in order. In spite of the phase space being only two-dimensional, this is just
enough to allow for inhomogeneous currents to flow (contrary to a one-dimensional phase space).
While the total magnetization has global fluctuations equal to those of a system in equilibrium at
$\beta$, it may be seen that the magnetization difference between the two spin groups is increased
with respect to its equilibrium counterpart in the presence of a current:
\begin{equation}
\langle (x_1-x_2)^2\rangle_{J\neq 0}-\langle (x_1-x_2)^2\rangle_{\hbox{\tiny eq},
J=0}=\frac{J^2}{4}\frac{2-\beta}{1-\beta}
\end{equation}
This provides an example of a nonequilibrium drive giving rise to an {\it increase} of fluctuations,
rather than to a {\it decrease} (as is usually noted, {\it e.g.} in driven lattice
gases~\cite{spohn,vanwijlandracz} and spin chains~\cite{eislerraczvanwijland}). We now turn to an analysis of the probability flow lines.

\subsection{Flow lines}
\begin{figure}[h]
  \begin{center}\epsfig{figure=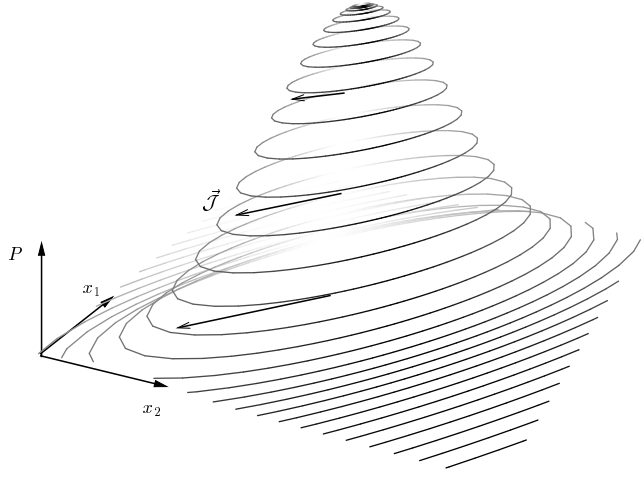,width=9cm}
  \end{center}
  \caption{The $(x_1,x_2)$ axes denote the deviations from the group 1 and 2 magnetizations. 
  Flow lines of the probability current $\vec{\cal J}$ are represented for different values
  of the constant $C$ in (\ref{ellipse}). The vertical axe is the probability $P(x_1,x_2)$,
  illustrating that the flow lines coincide with the isoprobability contours.
  On this figure, $\beta=0.8$ and $\eps=0.7$.}
  \label{flowline}
\end{figure}

In equilibrium, by definition, there is no
probability current, while in a NESS there are steady
(probability) currents. The flow lines, namely the set of points $(x_1,x_2)$ such that
$J_1(x_1,x_2)\dd x_2-J_2(x_1,x_2)\dd x_1=0$, in phase space turn out to be ellipses, as shown 
in Fig~(\ref{flowline}).
For $J\neq 0$ the flow lines are ellipses of equation
\begin{equation}\label{ellipse}
(1-\beta)\left(1+\frac{J^2}{4}\right)(x_1+x_2)^2+(x_1[1-J/2]-x_2[1+J/2])^2=C^2
\end{equation}
and they coincide with the isoprobability contours, a generic property of linear force driven systems. 
For an equilibrium system the flow lines can be seen to collapse onto a single point.
Interestingly, the shape of the flow lines can be used to infer properties of the steady-state
distribution: the departure from the ellipses will indicate deviations from linear forces in the
Fokker--Planck equation.

\subsection{Master equation and effective free energy}
We characterize a {\it state} of our system by the ``local" magnetizations
$M_\alpha=\sum_{j\in\alpha} \sigma_j$ and we denote (in this paragraph only) by
$m_\alpha=M_\alpha/N$. The steady-state solution to the master equation being denoted by 
$P_{\hbox{\tiny st}}(M_1,M_2)$, we define the effective free energy $f(m_1,m_2)$ by
\begin{equation}
f(m_1,m_2)=-\lim_{N\to\infty}\frac{\ln P_{\hbox{\tiny st}}(N m_1,Nm_2)}{N}
\end{equation}
We split $f$ into an entropic contribution $s(m_1,m_2)$ and an effective energy $e(m_1,m_2)$.
Whether in equilibrium or in a NESS, the entropic part is defined by the combinatoric factor for
the number of configurations with $M_1$ and $M_2$~:
\begin{eqnarray}
s(m_1,m_2)&=\frac{1}{N}
\ln {\frac{N}{2}\choose\frac{N+2M_1}{4} }{\frac{N}{2}\choose\frac{N+2M_2}{4}}
\nonumber\\
&=-\sum_\alpha\left[\frac{1+2m_\alpha}{4}\ln\frac{1+2m_\alpha}{4}
+\frac{1-2m_\alpha}{4}\ln\frac{1-2m_\alpha}{4}\right]
\end{eqnarray}
This fully defines $e=f+s$. In equilibrium we have that
$e_{\hbox{\tiny eq}}(m_1,m_2)=-\frac{\beta}{2}(m_1+m_2)^2$, and we wish to find how the nonequilibrium
drive modifies this result, namely what kind of effective interactions between the two groups of
spins it generates. Since $\sim\ee^{N(s-e)}$ is a stationary solution to the master equation
governed by the rates (\ref{rateE},\ref{rateO}), we find that $e(m_1,m_2)$ is a solution to
\begin{eqnarray}
\label{largedevenergy}
0&=(1-2m_1)\left(\ee^{-\beta_1(m_1+m_2)-2\p_1
e}-\ee^{+\beta_1(m_1+m_2)}\right)\nonumber\\
&+(1+2m_1)\left(\ee^{+\beta_1(m_1+m_2)+2\p_1 e}-\ee^{-\beta_1(m_1+m_2)}\right)\nonumber\\\
&+(1-2m_2)\left(\ee^{-\beta_2(m_1+m_2)-2\p_2 e}-\ee^{+\beta_2(m_1+m_2)}\right)\nonumber\\\
&+(1+2m_2)\left(\ee^{+\beta_2(m_1+m_2)+2\p_2 e}-\ee^{-\beta_2(m_1+m_2)}\right)
\end{eqnarray}
To first order in $\eps$ one may verify that 
\begin{eqnarray}
\label{eeff}
e(m_1,m_2)=-\frac 12 \beta m^2-\frac{\eps}{2-\beta} (m_1-m_2) h(m)+{\cal
O}(\eps^2),\,m=m_1+m_2
\end{eqnarray}
where the function $h$ is the solution to the following first order ordinary differential equation:
\begin{equation}
\left(m-\tanh\beta m\right)\frac{\dd h}{\dd m}+h(m)-(2-\beta)m=0,\;h(0)=0
\end{equation}
For instance, as $m\to 0$, 
\begin{equation}
h(m)= m-\frac{\beta^3}{3(1+3(1-\beta))}m^3+{\cal
O}(m^5)
\end{equation}
thus recovering the leading term of the high temperature Van Kampen expansion.  At this stage we
have completed our description of the steady-state properties. Note that the
structure of (\ref{eeff}) has flavors of the much more complex one found by
Derrida
{\it et al.}~\cite{derridalebowitzspeer} in the
framework of the asymmetric exclusion process for the effective free energy of a
given density profile.

\section{Energy current distribution}\label{groscalculs}
This section is devoted to determining the large deviation function of the time integrated energy
current.
\subsection{Modified master equation}
Let $Q(t)$ be the fluctuating energy current integrated over the time interval $[0,t]$ as defined in
(\ref{energycurrentdef}). We
are interested in $p(Q,t)$, the probability that $Q(t)=Q$ at time $t$ or in its generating function
$\hat{p}(\lambda,t)=\langle\ee^{-\lambda Q}\rangle$. It is possible to write
a master equation for 
\begin{equation}
P(M_1,M_2,Q,t)=\hbox{Prob}\{\sum_{j\in 1}\sigma_j=M_1\hbox{  and }\sum_{j\in 2}\sigma_j=M_2\hbox{  and
}Q(t)=Q\}
\end{equation}
Inserting the explicit expressions of the transition rates we arrive at
\begin{eqnarray}\label{masterQ}
\!\!\!\!\!\!\!\!\!\!\!\!\!\!\!\!\!\!\!\!\!\!\!\!\!\!\!\!\!\!\!\!\!\!\!\p_tP&=\frac{N/2+M_1+2}{N}\ee^{-\beta_1\frac{M+2}{N}}P(M_1+2,M_2,Q-2\frac{M+1}{N },t)-\frac{N/2-M_1}{N}\ee^{+\beta_1\frac{M}{N}}P(M_1,M_2,Q,t)\nonumber\\
\!\!\!\!\!\!\!\!\!\!\!\!\!\!\!\!\!\!\!\!\!\!\!\!\!\!\!\!\!\!\!\!\!\!\!&+\frac{N/2-M_1+2}{N}\ee^{+\beta_1\frac{M-2}{N}}P(M_1-2,M_2,Q+2\frac{M-1}{N },t)-\frac{N/2+M_1}{N}\ee^{-\beta_1\frac{M}{N}}P(M_1,M_2,Q,t)\nonumber\\
\!\!\!\!\!\!\!\!\!\!\!\!\!\!\!\!\!\!\!\!\!\!\!\!\!\!\!\!\!\!\!\!\!\!\!&+\frac{N/2+M_2+2}{N}\ee^{-\beta_2\frac{M+2}{N}}P(M_1,M_2+2,Q-2\frac{M+1}{N },t)-\frac{N/2-M_2}{N}\ee^{+\beta_2\frac{M}{N}}P(M_1,M_2,Q,t)\nonumber\\
\!\!\!\!\!\!\!\!\!\!\!\!\!\!\!\!\!\!\!\!\!\!\!\!\!\!\!\!\!\!\!\!\!\!\!&+\frac{N/2-M_2+2}{N}\ee^{+\beta_2\frac{M-2}{N}}P(M_1,M_2-2,Q+2\frac{M-1}{N },t)-\frac{N/2+M_2}{N}\ee^{-\beta_2\frac{M}{N}}P(M_1,M_2,Q,t)\nonumber\\
\!\!\!\!\!\!\!\!\!\!\!\!\!\!\!\!\!\!\!\!\!\!\!\!\!\!\!\!\!\!\!\!\!\!\!&
\end{eqnarray}
Going to the generating function
\begin{equation}
\hat{P}(M_1,M_2,\lambda,t)=\sum_Q\ee^{-\lambda Q}P(M_1,M_2,Q,t)
\end{equation}
and setting $|\Psi(\lambda,t)\rangle=\sum_{M_1,M_2}\hat{P}(M_1,M_2,\lambda,t)|M_1,M_2\rangle$, 
we may rewrite Eq.~(\ref{masterQ}) in the following form
\begin{equation}
\frac{\dd |\Psi(\lambda,t)\rangle}{\dd t}=-\hat{H}(\lambda)|\Psi(\lambda,t)\rangle
\end{equation}
where the operator $\hat{H }(\lambda)$ reads
\begin{equation}
\!\!\!\!\!\!\!\!\!\!\!\!\!\!\!\!\hat{H}(\lambda)=\sum_{j\in 1}\left(1-\sigma_j^x\ee^{+2\lambda\sigma_j^z
(M^z-\sigma_j^z)/N}\right)\ee^{-\beta_1 \sigma_j^z
M^z/N}+
\sum_{j\in 2}\left(1-\sigma_j^x\ee^{-2\lambda\sigma_j^z (M^z-\sigma_j^z)/N}\right)\ee^{-\beta_2 \sigma_j^z
M^z/N}
\end{equation}
The asymptotic behavior of 
\begin{equation}
\!\!\!\!\!\!\!\!\!\!\!\!\!\!\!\!\hat{p}(\lambda,t)=\sum_{M_1,M_2}\hat{P}(M_1,M_2,\lambda,t)=\langle {\bf
p}|\ee^{-\hat{H}(\lambda)t}|\Psi(0)\rangle,\;\langle {\bf
p}|=\sum_{M_1,M_2}\langle M_1,M_2|=\hbox{projection state}
\end{equation}
will be governed by the largest eigenvalue
$\mu(\lambda)$ of $-\hat{H}(\lambda)$ in the sense that
$\lim_{t\to\infty}\frac{1}{t}\ln\hat{p}(\lambda,t)=\mu(\lambda)$, which we now
set out to determine.
Before embarking into technicalities it is convenient, but by no means compulsory,
to perform a similitude transformation on $\hat{H}(\lambda)$
\begin{eqnarray}
\label{similitude}
\hat{H}_s(\lambda)&=\ee^{-\beta (M^{z})^2/4N}\hat{H}(\lambda)\ee^{+\beta
(M^{z})^2/4N}\nonumber\\&=
\sum_{j\in 1}\left(\ee^{-\beta_1 \sigma_j^z
M^z/N}-\sigma_j^x\ee^{+(2\lambda-\eps)\sigma_j^z
M^z/N-(2\lambda+\beta)/N}\right)\nonumber\\&+
\sum_{j\in 2}\left(\ee^{-\beta_2 \sigma_j^z
M^z/N}-\sigma_j^x\ee^{-(2\lambda-\eps)\sigma_j^z M^z/N+(2\lambda-\beta)/N}\right)
\end{eqnarray}
The transformation (\ref{similitude}) does not have the same effect as that conducted by
Ruijgrok and Tjon~\cite{ruijgroktjon} --it does not make the resulting operator
Hermitian-- but it serves the same practical purpose: calculations are performed
in a more convenient way where the system symmetries (upon exchanging the roles
of 1 and 2) are made obvious. In terms
of its symmetrized counterpart $\hat{H}_s(\lambda)$, we have that
\begin{equation}\label{GCmu}
\left(\hat{H}_s(\lambda)\right)^\dagger=\hat{H}_s(\eps-\lambda^*)
\end{equation}
An important consequence of symmetry (\ref{GCmu}) is that, for $\lambda$ real, both
$\hat{H}_s(\lambda)$ and $\hat{H}_s(\eps-\lambda)$ have the same spectrum, hence
\begin{equation}
\mu(\lambda)=\mu(\eps-\lambda)
\end{equation}
This is the Gallavotti--Cohen theorem. A direct consequence for the energy current large deviation
function $\pi(q)=\frac{1}{t}\ln p(Q=qt,t)$  is that
\begin{equation}
\pi(q)-\pi(-q)=\eps q
\end{equation}
where we have used that $\pi(q)={\hbox{max}}_\lambda\{\mu(\lambda)+\lambda q\}$. Another useful
consequence of (\ref{GCmu}) is that for $\lambda\in\frac{\eps}{2 }+i\mathbb{R}$,
$\hat{H}_s(\lambda)$ is Hermitian, which will justify diagonalization in that region of
the $\lambda$ complex plane.

\subsection{Mapping to a free boson problem}
We introduce, following Ruijgrok and Tjon~\cite{ruijgroktjon}, bosonic operators
$a_\alpha,a_\alpha^\dagger$ ($\alpha=1,2$) to
describe magnetizations 1 and 2 in the vicinity of the paramagnetic state:
\begin{equation}\label{holsteinprimakof}
M_\alpha^x=N/2-2a_\alpha^\dagger
a_\alpha,\;M_\alpha^y=-i\sqrt{N/2}(a_\alpha^\dagger-a_\alpha),\;M_\alpha^z=\sqrt{N/2}(a_\alpha^\dagger+a_\alpha)
\end{equation}
The relations (\ref{holsteinprimakof}) hold provided we are interested in states such that the
number operator $a_\alpha^\dagger a_\alpha$ remain of order
unity, that is much smaller than $\sqrt{N}$. In terms of these operators we find that
\begin{equation}
\hat{H}_s(\lambda)=\frac 12 \left(\begin{array}{cccc}a_1^\dagger&a_1&a_2^\dagger&a_2\end{array}\right)
\Gamma(\lambda)\left(\begin{array}{c}a_1^\dagger\\a_1\\a_2^\dagger\\a_2\end{array}\right)+\frac{1}{2}(\beta^2+4\lambda(\eps-\lambda))
\end{equation}
with
\begin{equation}
\Gamma=\left(\begin{array}{cccc}
Z-2\lambda& Z+2&Z&Z-(2\lambda-\eps)\\
Z+2&Z+2(\lambda-\eps)&Z+2\lambda-\eps&Z\\
Z&Z+2\lambda-\eps&Z+2\lambda&Z+2\\
Z-(2\lambda-\eps)&Z&Z+2&Z-2(\lambda-\eps)
\end{array}\right)
\end{equation}
where $Z=-\frac 12 \beta(2-\beta)+2\lambda(\eps-\lambda)$. For normalization
purposes~\cite{negeleorland} it is
necessary to define an auxiliary matrix $\tilde{\Gamma}$ built from $\Gamma$ by equating to zero in
the latter all matrix elements connecting two creation or two annihilation operators. It is then a simple matter~\cite{negeleorland} to determine not only the ground-state but also the spectrum of
$\hat{H}_s(\lambda)$. To do so we introduce the matrix $\Omega$ defined as
\begin{equation}
\Omega=\left(\begin{array}{cccc}
0&i\omega&0&0\\
-i\omega&0&0&0\\
0&0&0&i\omega\\
0&0&-i\omega&0
\end{array}\right)
\end{equation}
We must now evaluate the quantity
\begin{equation}
\mu(\lambda)=\frac{1}{2 }\int\frac{\dd \omega}{2\pi}
\ln\frac{\det(\tilde{\Gamma}+\Omega)}{\det({\Gamma}+\Omega)}-\frac{1}{2}(\beta^2+4\lambda(\eps-\lambda))
\end{equation}
Given that 
\begin{equation}
\det({\Gamma}+\Omega)=(\omega^2+\omega_+^2)(\omega^2+\omega_-^2)
\end{equation}
with
\begin{equation}
\omega_\pm(\lambda)=\sqrt{(2-\beta)^2+4\lambda(\eps-\lambda)}\pm\sqrt{\beta^2+4\lambda(\eps-\lambda)}
\end{equation}
and that similarly
\begin{equation}
\det(\tilde{{\Gamma}}+\Omega)=(\omega^2+\tilde{\omega}_+^2)(\omega^2+\tilde{\omega}_-^2)
\end{equation}
with
\begin{equation}
\tilde{\omega}_\pm(\lambda)=\frac 12\left((2-\beta)^2+4\lambda(\eps-\lambda)\pm
\sqrt{\left[(2-\beta)^2+4\lambda(\eps-\lambda)\right]\left[\beta^2+4\lambda(\eps-\lambda)\right]}
\right)\end{equation}
we arrive at the following result:
\begin{equation}
\mu(\lambda)=\frac{1}{2 }(\tilde{\omega}_++\tilde{\omega}_--\omega_+-\omega_-)-\frac{1}{2}(\beta^2+4\lambda(\eps-\lambda))
\end{equation}
which simplifies into
\begin{equation}\label{mulambda}
\mu(\lambda)=2-\beta-\sqrt{(2-\beta)^2+4\lambda(\eps-\lambda)}
\end{equation}
By the same token we obtain the spectrum of $\hat{H}(\lambda)$, whose
eigenvalues are given by
\begin{equation}
\hbox{Sp}(\hat{H}(\lambda))=\{\omega_+(\lambda)\ell+\omega_-(\lambda)\ell'-\mu(\lambda)\}_{\ell,\ell'\in\mathbb{N}}
\end{equation}
Specializing to $\lambda=0$ we obtain as a side result the spectrum of the master equation
evolution operator, whose slowest relaxation time is given by $\omega_+^{-1}(0)=(2(1-\beta))^{-1}$.
This again matches the results of \cite{ruijgroktjon}.\\

\begin{figure}[h]
\begin{center}\epsfig{figure=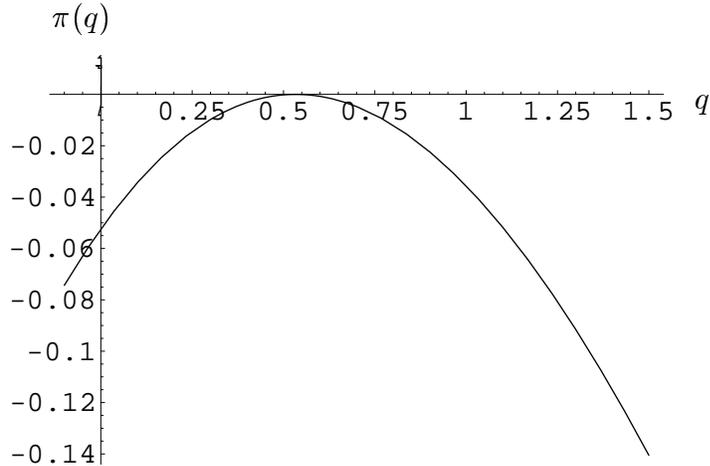,width=10cm}
\end{center}
\caption{This is the plot of the energy flux large deviation function $\pi(q)$ (vertical axis) as a function of $q$
(horizontal axis) at $\beta=0.5$ and $\eps=0.4$.}
\label{largepiq}\end{figure}

Given that $\pi(q)$ and $\mu(\lambda)$ are the Legendre transforms of each other we arrive at the
explicit form of the current large
deviation function $\pi(q)$
\begin{equation}
\pi(q)=\frac{\eps}{2}q+2-\beta-\sqrt{(2-\beta)^2+\eps^2}\sqrt{4+q^2}
\end{equation}
and it has the graph shown in Fig.\,\ref{largepiq}.

\subsection{Energy current in the low temperature phase}
\begin{figure}[h]
\begin{center}\epsfig{figure=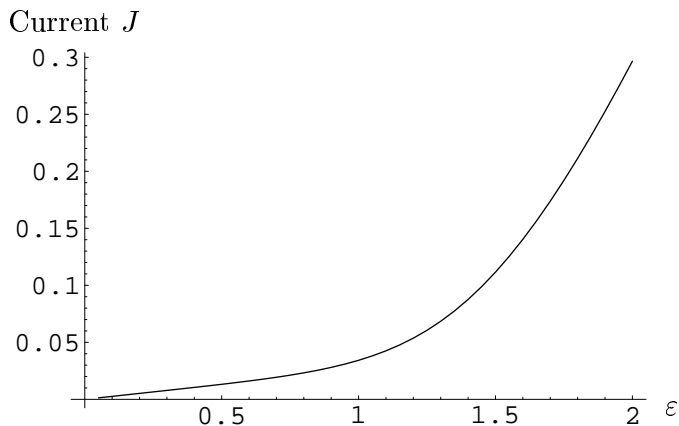,width=10cm}
\end{center}
\caption{Plot of the average energy current $J$ as a function of $\eps\in[0,\beta]$ at $\beta=2$ in the ordered phase.}
\label{figcourant}\end{figure}

Similar methods allowed us to express the generating function of the cumulants of $Q(t)$ in the low
temperature phase, at $\beta\geq 1$. The final result is
\begin{equation}
\mu(\lambda)=c_1+c_2-
\frac{1}{2}\left(\frac{\beta_1}{c_1}+\frac{\beta_2}{c_2}\right)-
\sqrt{\left[c_1+c_2-
\frac{1}{2}\left(\frac{\beta_1}{c_1}+\frac{\beta_2}{c_2}\right)\right]^2
+\frac{4}{c_1 c_2}\lambda(\eps-\lambda)}
\end{equation}
where $c_\alpha=\cosh\beta_\alpha (m_1+m_2)=1/\sqrt{1-4m_\alpha^2}$ and where $m_\alpha$ is the stationary solution of (\ref{autocoh}), and this is a function of $\beta$
and $\eps$. However the current is not the
order parameter of the phase transition, therefore nothing dramatic is expected to occur for
$\mu(\lambda)$ at $\beta=1$. An important difference with the high temperature result must be
emphasized: in the low temperature ordered regime the current is a nonlinear function of $\eps$, as
plotted in Fig.\,\ref{figcourant}.
The similar mathematical structure of $\mu(\lambda)$ in the high and low temperature phases seems to be generically related to
Langevin equations with linear forces~\cite{brunetderrida}.\\

\begin{figure}[h]
\begin{center}\epsfig{figure=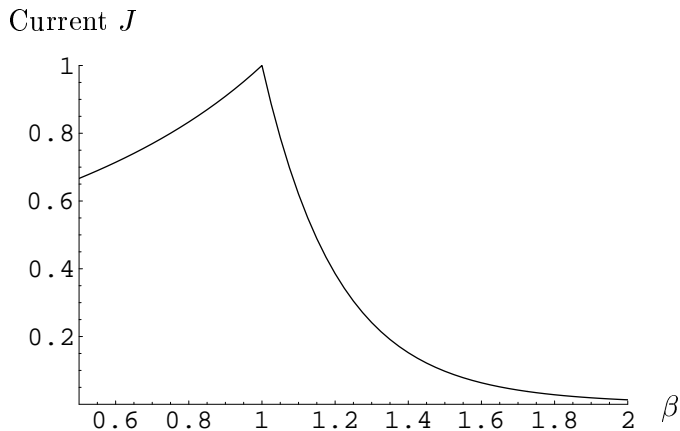,width=10cm}
\end{center}
\caption{Plot of the average energy current $J$ as a function of $\beta\in[0.5,2]$ at
$\eps=0.5$.} 
\label{figcourantbeta}\end{figure}

The energy current at fixed drive $\eps=0.5$ as a function of $\beta\in[0.5,2]$ represented in
Fig.\,\ref{figcourantbeta} shows that, from the disordered to the ordered phase, the current remains
finite and continuous, though it develops a cusp at the critical point $\beta=1$.

\subsection{Green-Kubo relations}
Exploiting the explicit formula (\ref{mulambda}) for $\mu(\lambda)$ we find, after differentiation
with respect to $\lambda$ once and twice, that
\begin{equation}
\frac{\langle Q\rangle}{t}=J=\frac{2\eps}{2-\beta},\;\; \frac{\langle
Q^2\rangle_c}{t}=\frac{4}{2-\beta}+\frac{4}{(2-\beta)^3}\eps^2
\end{equation}
Note that defining the diffusion coefficient $D(\beta)$ as the response to an external drive and
$\sigma(\beta)$ as the variance of the current fluctuations we find
\begin{equation}
D(\beta)=\frac{\p
J(\beta_1,\beta_2)}{\p\beta_1}\Big|_{\beta_1=\beta_2=\beta}=\frac{1}{2-\beta},\;\sigma(\beta)=\frac{\langle
Q^2\rangle_c}{t}\Big|_{\beta_1=\beta_2=\beta}
\end{equation}
With these expressions one may verify an integral formulation of the Green-Kubo relation
\begin{equation}
2\int_{\beta_2}^{\beta_1}\dd\beta\frac{D(\beta)}{\sigma(\beta)}=\eps
\end{equation}
Nevertheless the sole  knowledge of $D(\beta)$ and $\sigma(\beta)$  does not give access to the
full distribution $\mu(\lambda)$, as opposed to the cases studied by Bodineau and
Derrida~\cite{bodineauderrida} by means of an additivity principle or by Bertini {\it et
al.}~\cite{bertinidesolegabriellijonalasiniolandim1,bertinidesolegabriellijonalasiniolandim2} who
resorted to fluctuating
hydrodynamics~\cite{bertinidesolegabriellijonalasiniolandim1,bertinidesolegabriellijonalasiniolandim2}.
In order the latter approaches to hold, the typical current must scale to zero with the system size at fixed
(intensive) external field. This the second example, aside from the extensively studied Asymmetric
Exclusion Process~\cite{derridalebowitz}, of an interacting system, albeit mean-field, in which the
entropy (or energy) current can be computed exactly, with the additional property in our case that
the Gallavotti--Cohen theorem is fulfilled, hence generalized Green--Kubo relations as well.

\section{Final comments}\label{finalcomments}
We have been able to present explicit and exact results for the steady-state of a system made of interacting spins driven far from
equilibrium by heat baths at different temperatures. The system described exhibits a
ferromagnetic-to-paramagnetic phase transition. The simplicity of some of our results, like
Gaussian fluctuations for the magnetizations, are admittedly an artifact of our infinite-range,
mean-field, model. Nevertheless, due to easier mathematics, we have been able to precisely describe
the probability flow lines, ellipses in a
two-dimensional phase space. Other concepts arising within the framework of dynamical systems
theory,
like that of topological pressure, once transposed to our model, might equally lend themselves to analytical
approaches.\\

Our other result of interest concerns the computation of an energy current distribution for a system
far from equilibrium, that cannot be described by fluctuating hydrodynamics, although it falls
within the scope of the Gallavotti--Cohen theorem for Markov processes~\cite{lebowitzspohn}. To our
knowledge, this is the first one of this sort. It is only a first step towards the desirable, but
remote, goal of characterizing stationary systems driven very far from equilibrium.\\

Among the prospects uncovered by the present work, we mention the extension of the urn
model of Bena {\it et al.}~\cite{benacoppexdrozlipowski} analyzed in the light of
Greenblatt and Lebowitz' comment~\cite{greenblattlebowitz}. In spite of being genuinely
nonequilibrium, our model will most probably not display any surprises as far as the Yang-Lee zeros of
the partition function are concerned, simply because the phase transition that takes place at
$\beta=1$ is akin to its equilibrium
counterpart (and belongs to the same universality class). However the urn model may be very well defined
for negative temperatures. Preliminary studies indicate that such rates open the door to limit cycles
and chaotic behavior that we shall explore in future studies.\\ 

\noindent {\bf Acknowledgement}: This research has been
partially supported by the Hungarian Academy of Sciences (Grant
No. OTKA T043734). It is a pleasure for the authors to thank Henk Hilhorst, C\'ecile
Appert and Bernard Derrida for their comments during the preparation of this work.

\section*{Appendix A: Fokker--Planck equation for $n$ heat baths}

In the disordered phase, the Ising system is split in $n$ parts of
magnetization fluctuation $x_\alpha$ ($1\leq \alpha\leq n$), each of them
equilibrated with a bath at inverse temperature
$\beta_\alpha=\beta+\eps_\alpha$, where $\beta$ is the average of the
$\beta_\alpha$'s. The Fokker--Planck stationary equation for the whole
system can be written $\partial_\alpha {\cal J}_\alpha = 0$ (we will use summation
convention over repeated indices). The ${\cal J}_\alpha$'s are the components of
the probability current, of expression~:
\begin{eqnarray}
  {\cal J}_\alpha    & = -\partial_\alpha P + f_{\alpha\gamma}x_\gamma P \\
  f_{\alpha\gamma} & = - n \delta_{\alpha\gamma} + \beta_\alpha \    \label{eqn:force}
\end{eqnarray}

This is a special case of a Fokker-Planck equation with linear forces $f_{\alpha\gamma}x_\gamma$.
As $f$ is a definite negative matrix, we know from general
considerations~\cite{vankampen} that its stationary state is Gaussian of
covariance matrix~:
\begin{equation} \label{eqn:cov_mat}
 \int_0^\infty dt\
 e^{t f} e^{t f^T}\ 
\end{equation}
This matrix however proves difficult to be explicitated in the
general case. Here we will restrict our analysis to the case where $f$ can be
diagonalized, as in our system, and use an indirect method to compute the result
of (\ref{eqn:cov_mat}). First note that
if $f$ is symmetric, setting $U(x_1,\ldots,x_n)=-\frac{1}{2}x^Tfx$, we have
$f_{\alpha\gamma}x_\gamma=-\partial_\alpha U$. In other words, the forces applied on the system
are conservative and deriving from the potential $U$. In that case, we
immediately find the stationary solution of the Fokker--Planck equation
by requiring ${\cal J}_\alpha=0$~: this is, as expected, the Gibbs--Boltzman distribution
$P(x)\sim \ee^{-U(x)}$, which corresponds to an equilibrium situation,
and the defining absence of a probability current.\\

For $f$ diagonalizable but not symmetric, no solution
can be found imposing ${\cal J}_\alpha=0$.  Our aim is to parallel the resolution of
the equilibrium case, by transforming the Fokker--Planck equation into some new
equation $\partial_\alpha {\cal J}_\alpha^\prime=0$ whose solution can be found by
requiring ${\cal J}_\alpha'=0$.  We first perform the change of
variable $y=bx$, where $b$ is a matrix such that $b f b^{-1} = {\rm Diag
}(\lambda_1,\ldots,\lambda_n)$.  The Fokker--Planck equation takes the form~:
\begin{equation}\label{eqn:FP_y}
 - b_{\alpha\alpha'} b_{\gamma\alpha'} \partial_\alpha \partial_\gamma P + \partial_\alpha
 (\lambda_\alpha y_\alpha P) =0\ 
\end{equation}
where now $\partial_\alpha=\partial/\partial y_\alpha$. The force term is 
symmetric, and can be formally written as deriving from some potential.
There are many ways to split the first term of~(\ref{eqn:FP_y})	so as
to write it as a divergence. Let $A_{\alpha \gamma}$ be arbitrary constants.
Writing, for $\alpha<\gamma$:
\begin{equation}
  b_{\alpha\alpha'} b_{\gamma\alpha'} \partial_\alpha \partial_\gamma P =
   \partial_\alpha ( A_{\alpha\gamma}     b_{\alpha\alpha'} b_{\gamma\alpha'}\partial_\gamma P  )+
   \partial_\gamma ( (1-A_{\alpha\gamma}) b_{\alpha\alpha'} b_{\gamma\alpha'}\partial_\alpha P  ) \ ,
\end{equation}
the FP equation now takes the announced form
$\partial_\alpha {\cal J }_\alpha^{\prime}=0$, and requiring ${\cal J }_\alpha^{\prime}=0$
means (defining $\vec b_\alpha\cdot \vec b_\gamma=b_{\alpha\alpha'} b_{\gamma\alpha'}$)~:
\begin{eqnarray}
\fl
\underbrace{
  \left(
  \begin{array}{ccccc}
    \vec b_1^2 &  & &  \left(2A_{\alpha\gamma} \vec b_\alpha\cdot \vec b_\gamma\right) \\
   &  \vec b_2^2 \\
   & & \ddots \\
   &\left(2(1-A_{\gamma\alpha})\vec b_\alpha\cdot \vec b_\gamma\right) &
       & & \vec b_n^2 \\
   \end{array} \right) }_L
   \; \vec\nabla P =  
%
\underbrace{
{\rm Diag }(\lambda_1,\ldots,\lambda_n) }_\Lambda
\vec y
\end{eqnarray}
This equation has a solution if, and only if $\Lambda^{-1} L$ is symmetric,
which occurs when:
\begin{equation}
  A_{\alpha\gamma} = \frac{\lambda_\gamma}{\lambda_\alpha+\lambda_\gamma}
\end{equation}
In terms of the original $x$ variables, we finally get the following expression~:
\begin{equation} \label{eqn:sol_FP_general}
  P(x) = \exp\left(-\frac{1}{2}x^T b^T M_y^{-1} b\: x\right) 
  \qquad {\rm with } \qquad
  (M_y)_{\alpha\gamma}= \frac{2 \vec b_\alpha\cdot\vec b_\gamma}{\lambda_\alpha+\lambda_\gamma}
\end{equation}
in which we can read the result of~(\ref{eqn:cov_mat}) in the general case.
When applying this result to the $n$ baths Ising problem, we find~:
\begin{equation}
  P(x) =  \exp\left(-\frac{1}{2}x^T M^{-1} x\right)
 \
 {\rm with :}
 \
 M_{\alpha\gamma}
  = \frac{1}{n}\: \delta_{\alpha\gamma} \ +
 \frac{1}{n^2} \left(-\frac{1-\beta}{2-\beta}+
\frac{(1+\eps_\alpha)(1+\eps_\gamma)}{(1-\beta)(2-\beta)} \right)
\end{equation}
The method above can be generalized to FP equations whose diffusion term
is $\Gamma_{\alpha\gamma} \partial_\alpha \partial_\gamma P$ (as, for instance, in the low-temperature
phase of our Ising model). The final result takes the form (\ref{eqn:sol_FP_general})
with $ (M_y)_{\alpha\gamma}= 2 \Gamma_{\alpha'\gamma'}
b_{\alpha\alpha'}b_{\gamma\gamma'}/(\lambda_\alpha+\lambda_\gamma)$.

\section*{Appendix B: entropy current distribution function for $n$ heat baths}

As in appendix A, we consider an Ising system split in $n$ parts equilibrated
with baths at inverse temperature $\beta_\alpha$ with $1\leq \alpha\leq n$. The Lebowitz
and Spohn~\cite{lebowitzspohn} integrated entropy current~(\ref{entropycurrent})
cannot be
decomposed as in~(\ref{entropycurrentdecomposition}). We will thus study the distribution
of~${\cal Q}_S(t)$.
As in section~\ref{groscalculs} the corresponding function $\mu(\lambda)$ is given
by the largest eigenvalue of some operator~$-\hat{H}(\lambda)$, which reads~:
\begin{equation} \label{eqn:H_L_nbaths}
 \hat{H} (\lambda) = \sum_{\alpha=1}^n \sum_{j\in \alpha}
 \left(
  \ee^{-\sigma_j^z\beta_\alpha M_j^z / N}-\sigma_j^x 
  \ee^{(2\lambda-1)\beta_\alpha \sigma_j^z M_j^z / N}
\right) e^{-\beta_\alpha/N}\ 
\end{equation}
where we defined the operator $M_j^z=\sum_{i\neq j} \sigma_i^z$
verifying $[M_j^z,\sigma_j^x]=0$. As expected on general
grounds~\cite{kurchan,lebowitzspohn}, this operator possesses the
Gallavoti--Cohen symmetry~: for $\lambda$ real,
$(\hat{H}(\lambda))^\dagger=\hat{H}(1-\lambda)$. There is no need here to
perform a symmetrization analogous to (\ref{similitude}) because we directly focus on the {\it total}
entropy current ${\cal Q}_S(t)$.

For the sake of simplicity, we will now confine our analysis to the
disordered phase ($\beta<1$). In that phase the magnetizations can be
expanded up to order ${\cal O}(N^{-1/2})$ in terms of bosonic
operators $a_\alpha,a_\alpha^\dagger$~:
\begin{equation}
  M^x_\alpha = N/n - 2 a^\dagger_\alpha a_\alpha, \ \
  M^y_\alpha = -i \sqrt{N/n}(a^\dagger_\alpha -a_\alpha), \ \
  M^z_\alpha =  \sqrt{N/n}(a^\dagger_\alpha +a_\alpha), \ \
\end{equation}
yielding the following expression, still bearing the Gallavotti-Cohen symmetry~:
\begin{eqnarray}
 \hat{H} (\lambda) =& \ 
  \frac{2}{n}\lambda(1-\lambda)(\beta^2+\sigma^2)
  \Big(\sum_\alpha (a_\alpha^\dagger+a_\alpha)\Big)^2 \nonumber\\
  &\ \ -\frac{2}{n}\sum_{\alpha,\,\gamma} \beta_\gamma
  \big((1-\lambda)a_\gamma^\dagger+\lambda a_\gamma\big) (a_\alpha^\dagger+a_\alpha)
  +2\sum_\alpha a_\alpha^\dagger a_\alpha + 2 \lambda \beta  \nonumber\\
\end{eqnarray}
where we define and recall that~:
\begin{equation}
  \sigma^2 = \frac{1}{n} \sum_\alpha \eps_\alpha^2\ ,  \ \
  \beta    = \frac{1}{n} \sum_\alpha \beta_\alpha
\end{equation}
As in section~\ref{groscalculs}, we find $\mu(\lambda)$ is given by~:
\begin{equation}\label{eqn:mu_lambda_det}
 \mu(\lambda) = \frac{1}{2}\int \frac{\dd\omega}{2\pi}\:
 \ln\:\frac{\det (\tilde\Gamma+\Omega)}{\det(\Gamma+\Omega)}
-
 2 \lambda (1-\lambda) (\beta^2+\sigma^2) \ 
\end{equation}
where $\Gamma$, $\tilde\Gamma$ and $\Omega$ are $2n\times 2n$ matrices defined
by blocks in the following way~:
\begin{equation}
 \Gamma(\lambda) = 2 \:
 \left(
 \begin{array} {cc}
   A & C^\dagger \\
   C & A'
 \end{array} \right), \ \
 \tilde\Gamma  =  2 \:
 \left(
 \begin{array} {cc}
   0      & C^\dagger \\
   C & 0
 \end{array} \right),
\\
 \Omega  =
 \left(
 \begin{array} {cc}
   0                    & i \omega {\rm Id} \\
   -i \omega {\rm Id} & 0
 \end{array} \right)
\end{equation}
with elements~:
\begin{eqnarray} \label{eqn:def_alpha_beta}
 A_{\alpha\gamma} &=  \frac{2}{n}\lambda(1-\lambda)\,(\beta^2+\sigma^2)
                -\frac{\lambda}{n}\:(\beta_\alpha+\beta_\gamma) \\
 A'_{\alpha\gamma}&=  \frac{2}{n}\lambda(1-\lambda)\,(\beta^2+\sigma^2)
                -\frac{1-\lambda}{n}\:(\beta_\alpha+\beta_\gamma) \\
C_{\alpha\gamma}  &=  \frac{2}{n}\lambda(1-\lambda)\,(\beta^2+\sigma^2)
                -\frac{1}{n}\:\big((1-\lambda)\beta_\alpha+\lambda\beta_\gamma\big)
                +\delta_{\alpha\gamma}
\end{eqnarray}
Tedious yet straigtforward calculations lead to the following expressions~:
\begin{eqnarray}
  \det (\Gamma+\Omega) &= (-1)^n(\omega^2+4)^{n-2}
  \left(
    \omega^2+\omega_+^2
  \right)
  \left(
    \omega^2+\omega_-^2
  \right) \\
  \det(\tilde\Gamma+\Omega) &= (-1)^n (\omega^2+4)^{n-2}
  \left(
    \omega^2+\tilde\omega_+^2
  \right)
  \left(
    \omega^2+\tilde\omega_-^2
  \right) \\
  \omega_\pm &= \sqrt{(2-\beta)^2+4\lambda(1-\lambda) \sigma^2}
            \pm\sqrt{\beta^2+4\lambda(1-\lambda) \sigma^2} \\
 \tilde\omega_\pm &=
  2-\beta+2\lambda(1-\lambda)(\beta^2+\sigma^2) \pm \nonumber \\ & \qquad
  \sqrt{4\lambda(1-\lambda)\sigma^2+ 
      \big(2\lambda(1-\lambda)(\beta^2+\sigma^2)-\beta\big)^2} 
\end{eqnarray}
The expression of  $\mu(\lambda)$ simplifies into:
\begin{equation} \label{eqn:mu_nbaths}
  \mu(\lambda) = 2-\beta-\sqrt{(2-\beta)^2+4\lambda(1-\lambda)\sigma^2} 
\end{equation}
We again obtain the full spectrum of the evolution operator:
\begin{equation} 
   {\rm Sp } \ \hat{H}(\lambda) =
   \{2k+\omega_+(\lambda) \ell+\omega_-(\lambda)\ell' - \mu(\lambda) 
     \}_{k,\ell,\ell'\in\mathbb N}
\end{equation}
whose degeneracy is $k+n-3 \choose k $. In the high temperature phase,
the modes of the evolution operator split into two groups: two modes
are similar to the $n=2$ bath case, and $n-2$ modes relax with constant rate $2$. The slowest relaxation time
of the master equation evolution operator is the same as that found for $n=2$. The components of
the probability current along the corresponding $n-2$ directions cancel in the NESS, thus leaving
the probability flow lines as ellipses, as is the case for $n=2$.

Finally, we can remark that the result~(\ref{eqn:mu_nbaths}) for
$\mu(\lambda)$ could also have been found by studying the set of Langevin
equations associated to the Fokker--Planck equation of the system. However,
the Langevin
formalism  does not render the Gallavotti--Cohen symmetry explicit at intermediate steps of the calculations, as opposed to the
operator approach chosen in Sec.~4.

\end{document}